\documentclass[12pt]{article}
\usepackage{avant}
\usepackage{psfrag,color}
\newcommand{\e}{\mbox{\large e}}
\newcommand{\pr}{^{\prime}}

\newcommand{\deltad}[2]{\delta_{{#1},{#2}}}

\newcommand{\ctgh} {\rm ctgh \,}

\newcommand{\G}[2]{{\cal G}_{#1}^{#2}}

\newcommand{\J}[2]{{\cal J}_{#1}^{#2}}

\newcommand{\Z}[2]{{\cal Z}_{#1}^{#2}}

\newcommand{\multisum}[2]{\sum_{ \{ {#1}_{#2} \} } \;\;}

\newcommand{\cutsum}[2] { {\Lambda {#1} \over {#2} N }  \; 
                           \sum_{|\omega| < \Lambda}}

\begin{document}
\baselineskip=15pt
\begin{Large}
\noindent
{\bf Duality $\,$  in  the  $\,$ Quantum  $\,$ Dissipative  $\,$ Villain  
        Model  and application  to  Mesoscopic  Josephson  Junction  Circuits}  
\end{Large}

\vspace{5mm}
\noindent
G. Falci
\footnote{Istituto di Fisica, Universit\'a di Catania \& INFM,
         viale A. Doria 6, 95125 Catania (Italy)}
and U. Weiss
\footnote{ II. Institut f\"ur Theoretische Physik, 
   Universit\"at Stuttgart, 70550 Stuttgart (Germany)}

\vspace{5mm}
\hspace{1.4cm}
\begin{minipage}{11.5cm}
{\bf Abstract} 

\noindent
We study exact self duality in the model of a Brownian 
particle in a washboard (WB) potential
which describes a Josephson Junction (JJ) coupled to an 
environment, for arbitrary temperature and arbitrary 
form of the spectral density of the environment.

To this end we introduce the Quantum Dissipative Villain Model 
(QDVM), which models tunneling of a degree of freedom
coupled to a linear quantum environment through an infinite set of 
states.
We derive general exact mappings on various dual discrete 
representations
(one-dimensional Coulomb gases or surface roughening models)
which are exactly self-dual.
Then we show how the QDVM maps exactly onto the WB model
and use duality relations to calculate the leading terms of 
the total impedance of a JJ circuit, for general frequency dependence
of the spectral density of the environment and arbitrary temperature.

\vspace{4mm}
\end{minipage}

\vspace{15mm}
\noindent
\begin{minipage}{7cm}
{\bf Correspondent author} \\
Dr. Giuseppe Falci \\
Istituto di Fisica, Universit\'a di Catania \\ 
viale A. Doria 6, 95125 Catania (Italy) \\
        tel. : +39 095 7372435 \\
         fax: +39 095 333231 \\
        e-mail: gfalci@cdc.unict.it
\end{minipage}

\newpage
\baselineskip=22pt
\section{\large{INTRODUCTION}}
\label{sec:Introduction}
Tunneling in quantum dissipative systems
can be described in terms of
one (or few) special degree of freedom coupled to an incoherent 
environment of quantum oscillators~\cite{kn:Weiss-98}.
If the special variable has an underlying 
discrete character the model may have various 
discrete dual representations, 
each corresponding to
specific dressed excitations.

As an example of underlying discrete character of a continuous 
degree of freedom we may consider 
a Brownian particle in a double-well potential~\cite{kn:Weiss-98},
%\cite{kn:Weiss-Grabert}
a 
model which also describes a mesoscopic 
SQUID~\cite{kn:Leggett}.
In some region of the parameter space the relevant physics is well 
represented by
a two-state approximation, each state being  
related to a minimum of the potential.

In this paper we will focus on the WB model~\cite{kn:Caldeira-Leggett}
which describes the Brownian motion of a quantum particle in a tilted 
washboard potential and the quantum dynamics of mesoscopic 
JJ circuits~\cite{kn:Caldeira-Leggett,kn:revs},
%of Luttinger Liquids~\cite{kn:Kane-Fisher}
%with barriers and of 
%edge states in the Fractional Quantum Hall Effect regime~\cite{kn:Fisher}. 
For JJs the above special variable is 
the phase $\varphi$ across the junction (see fig.1)
%(see fig.\ref{fig:JJcircuit}),
a collective degree
which interacts whith all the microscopic variables entering
the problem (quasiparticles in the electrodes~\cite{kn:revs},
excitations of the circuit~\cite{kn:Ingold-Nazarov})
modeled by an environment with given spectral density.
The partition function is 
\begin{eqnarray}
  \label{eq:Z-WB}
 && \!\!\!  \!\!\!  \!\!\!\!\!\!  \!\!\! {\cal Z}({\cal I}_x) =
  \int {\cal D}\varphi_{\tau} \,\,
                     \mbox{\large e}^{\,
- \,{1 \over 2 }  \, \int_{0}^{\beta} d\tau  d\tau^{\prime} \,
          \varphi(\tau)   \, {\cal A}^{0}(\tau - \tau^{\prime}) 
         \, \varphi(\tau^{\prime}) \,+ \, \int_{0}^{\beta} d\tau  \, 
                         {\cal I}_x(\tau)  \varphi(\tau) }   \;
  \mbox{\large e}^{\,  \int_{0}^{\beta} d\tau  \,
                         V  \, \cos \varphi(\tau)}
\\
  \label{eq:Act-WB}
&& {\cal A}^{0}(\omega) \;=\; m \omega^2 - {\alpha(\omega) \over 2 }
\qquad \longleftrightarrow \qquad  { |\omega| \over 2 \pi}  \;
{R_Q \over Z_T(-i |\omega| )}  \; ,
%=m \omega^2  - { \alpha(\omega) \over 2 }
\end{eqnarray}
where $m$ is the mass and $\alpha(\omega)$ is the damping kernel for the 
Brownian particle~\cite{kn:Weiss-98}. They correspond to $C$ and 
$Z(\omega)$ in the JJ circuit of fig.1,
%fig.\ref{fig:JJcircuit},
 $Z_T(\omega)$ being the impedance seen by the 
junction~\cite{kn:Leggett-84,kn:Falci-Bubanja-Schoen,kn:Ingold-Nazarov},
$Z_T(\omega) = \left( 1 / Z(\omega) + i \omega C \right)^{-1}$.

The WB model accounts for a variety of effects 
(eg. Josephson effect, phase diffusion, macroscopic quantum phenomena, 
Coulomb blockade, Bloch oscillations)  
in different regimes~\cite{kn:revs},
due to its ``dual'' structure.
Indeed $\varphi$ and the charge $Q$ at the 
junction are conjugate quantum variables,
$\left[\hbar \varphi / 2e , Q \right]= i \hbar$ and
both have an underlying discrete character
(it comes from the existence of minima in the WB potential
for $\varphi$, whereas for
$Q$ it is related to the fact that $Q$ varies 
in units of $2e$ due to Cooper pair tunneling). 
As a result pairs of ``dual'' effects are observed in JJ 
circuits~\cite{kn:revs},
for instance the Josephson effect (a finite current flowing at zero 
voltage, corresponding to classical dynamics of $\varphi$) 
and the
Coulomb blockade (finite voltage drop with 
zero current, corresponding to classical  stochastic dynamics of $Q$).
It is tempting to think that ``quantum duality'' implies a phase transition 
triggered by the ratio $V/E_C$, $E_C=e^2/2C$ being the 
charging energy~\cite{kn:revs}, but this is wrong: 
the low-frequency response of a 
{\em single} 
junction device in a circuit is determined solely by the properties of 
the coupling with the environment~\cite{kn:Schmid}. 
For instance, for an ohmic environment 
the critical parameter is 
$\alpha = R_Q / Z(\omega=0)$ ($R_Q = h/4 e^2 \approx 6.5 \, k\Omega$ is
the quantum of resistance): Coulomb blockade is possible if 
$\alpha \ll 1$ (effectively current biased junction), whereas for
$\alpha \gg 1$ (effectively voltage biased junction) 
the dynamics of $\varphi$ is semiclassical.

In general, collective excitations and duality 
are more involved than the Kramers-Wannier 
$\varphi$-$Q$ duality because they depend mainly on the 
environment~\cite{kn:Schmid,kn:Fisher-Zwerger}.
Duality in the WB model with ohmic dissipation at $T=0$
was studied first by Schmid~\cite{kn:Schmid} 
who found a low-frequency approximate self-duality
with $\alpha \leftrightarrow 1/\alpha$.  
A dual mapping onto a one-dimensional 
Tight-Binding model was then introduced~\cite{kn:Fisher-Zwerger} for
arbitrary temperature, which was then used by 
Sassetti et al.~\cite{kn:Sassetti} to map into each other the sub-ohmic
and the super-ohmic environment. Very recently it was shown 
that the Schmid self-duality is exact at $T=0$ for
a strictly ohmic environment~\cite{kn:Fendley-Saleur}
the $m,C \to 0$ and $Z_T(\omega)=R$ limit of eq.(\ref{eq:Act-WB}),  
where the problem is exactly solvable~\cite{kn:exact}. 

In this paper we study the discrete representations 
and duality for the WB model. We 
introduce in sec.\ref{sec:QDVM} the QDVM and its discrete 
dual and self-dual representations, surface roughening and Coulomb 
gas models~\cite{kn:Chui-Weeks-82}.
The QDVM, besides illustrating in a simple but exact mathematical framework 
how duality works in connection with discreteness and dissipation,
can be reduced exactly
to the WB model with a given environment (sec.\ref{sec:WB}). 
thus, also the WB model has an 
{\em exact self-dual structure for arbitrary environment and temperature}. 
As an application we finally calculate the impedance of a 
Josephson circuit for low frequencies.

\section{\large{THE QDVM}}
\label{sec:QDVM}
\subsection{The model}
\label{sec:model}
The QDVM is defined on the discretized (slice $1/\Lambda$) imaginary time axis as 
\begin{eqnarray}
  \label{eq:Z-Villain}
  {\cal Z}({\cal J}) &=& 
  \int_{-\infty}^{\infty}  \left[ \prod_{\tau} d \varphi_{\tau} \right] \;\;
                     \mbox{\large e}^{-\,
{1 \over 2 \Lambda^2}  \; \sum_{\tau,\tau^{\prime}} \;  
          \varphi_{\tau}   \; {\cal A}_{\tau  \tau^{\prime}}^{0} 
         \; \varphi_{\tau^{\prime}} + \; {1 \over \Lambda} \; \sum_{\tau}  \; 
                         {\cal J}_{\tau}  \varphi_{\tau} }
\nonumber\\
&& \hskip40mm \cdot \quad
\sum_{\{m_\tau \}}  \mbox{\large e}^{-\, {1 \over 2 \Lambda}  \; \sum_{\tau}  
\; V  \; \left( \varphi_{\tau}
                                    - 2 \pi m_{\tau} \right)^2} \; .
\end{eqnarray}
The discrete variable $m_{\tau}$ is coupled 
to $\varphi_{\tau}$ which undergoes gaussian 
fluctuations tuned by the kernel ${\cal A}_{\tau  \tau^{\prime}}^{0}$,   
${\cal J}_{\tau}$ being the source. 
We can interpret eq.(\ref{eq:Z-Villain}) as the discretized path integral for
$ \varphi(\tau) - 2 \pi m(\tau)$, 
a continuous variable with an underlying discrete character 
(see fig.2a)
%(see fig.\ref{fig:discretex}a), 
coupled to an environment.

By comparing eq.(\ref{eq:Z-Villain})
with eq.(\ref{eq:Z-WB}) 
it is apparent that the QDVM can be obtained from
the discretized version of the WB model by
using the Villain approximation~\cite{kn:Villain},
so we can identify the kernel ${\cal A}^{0}_{\tau \, \tau^{\prime}}$. 
The Villain approximation has been successfully 
used to study 
mesoscopic JJ arrays~\cite{kn:arrays}, although it is not valid 
in the continuum limit $V/\Lambda \to 0$.
Notice however that here we use a strategy different from 
ref.~\cite{kn:arrays}: 
we first discuss the dual properties of
the QDVM and then we show how to obtain {\em exactly} from it the WB model,
in the limit $V/\Lambda \to 0$.

\subsection{Discrete representations of the QDVM}
\label{sec:discrete-repr}
\paragraph{Surface roughening representation}
Starting from 
eq.(\ref{eq:Z-Villain}), we can eliminate $\{\varphi_{\tau} \}$
and obtain a surface roughening model~\cite{kn:Chui-Weeks-82} 
on the imaginary time axis, where $m_{\tau}$
is the surface height in $[\tau,\tau+1/\Lambda]$ (see fig.2b).
%(see fig.\ref{fig:discretex}b). 
Each element $m_{\tau}$ of 
the surface interacts with all the other elements.

\paragraph{Coulomb gas representation I: $e$ charges}
We now introduce the charges $e_{\tau} = m_{\tau+1/\Lambda} - m_{\tau}$ and
perform a discrete double integration of the interaction to
obtain the generating functional of the QDVM in the $e$-representation
\begin{eqnarray}
\label{eq:Z-e-repr}
%--------------------------------------------------------------------
\Z{}{}(\J{}{}) &=& \Z{}{V}(\J{}{}) \cdot \Z{}{(e)}(\J{}{(e)})
\nonumber\\
\Z{}{(e)}(\J{}{(e)}) &=&  \multisum{e}{\tau} 
                   \e^{-{1 \over 2} \; 
                     \sum_{\tau,\tau \pr =  0}^{N-1} \;
                   e_{\tau}  \; \Delta_{\tau \, \tau \pr} \;
                   e_{\tau \pr}
                       + \; \sum_{\tau}  \; 
                         \J{\tau}{(e)} e_{\tau} } \; .
%--------------------------------------------------------------------
\end{eqnarray}
This is a set of interacting 
($\Delta_{\tau \, \tau \pr}$) integer charges 
$e_{\tau} \in ]-\infty, \infty[$ (see fig.2c).
%(see fig.\ref{fig:discretex}c). 
$\Z{}{V}(\J{}{})$ is the generating functional
for the damped harmonic oscillator~\cite{kn:Weiss-98}, and
the interaction is given by
\begin{eqnarray}
%--------------------------------------------------------------------
  \label{eq:kernel-e-repr}
  \Delta_{\tau} =  
%~~~~~~
  \cutsum{}{} \, 
                 {
\left\{ 4 \pi^2 m - 
                         {4 \pi^2 \, \alpha(\omega) / 2 \omega^2 }  
                       \right\} \; 
                V    \over 
                   m \omega^2 - \alpha(\omega)/2 + V
                   } \;\; \e^{-i \omega \tau} \; .
%--------------------------------------------------------------------
\nonumber
\end{eqnarray}
%For ohmic dissipation and large $\tau$,  
%$\Delta_{\tau} \sim 2 \alpha \ln \tau$. 
The source 
$\J{\tau}{(e)}$ can be obtained
from the source $\J{\tau}{}$ of the
QDVM~\cite{kn:Falci-Weiss}.

If we classify the configurations in 
eq.(\ref{eq:Z-e-repr}) according to the number of charges 
of each species we obtain the Coulomb gas representation 
\begin{eqnarray}
\label{eq:Z-e-repr-all}
\Z{}{(e)}(\J{}{(e)}) = 
           \multisum{\cal M}{\bar{e}}  \!\!\!
           \left[  \prod_{\bar{e}=-\infty}^{\infty} 
              \left( {\cal M}_{\bar{e}}! \right)^{-1} \right]
           \sum_{\{\tau_l \}} %_{\neq}}
                   \e^{-{1 \over 2} 
                     \sum_{l,l \pr =  1}^{{\cal M}_{T}} 
                   e_{l}   \Delta_{\tau_l  \tau_{l \pr}}
                    e_{l \pr}
                       +  \sum_{l}  
                         \J{\tau_l}{(e)}  e_{l} } \; .
\end{eqnarray}
Here each term of the summation is labeled by a set 
$\{ {\cal M}_{\bar{e}} \}$ where  ${\cal M}_{\bar{e}}$ is
the number of charges $\bar{e} \neq 0$ present. The 
charges are numbered with the index $l=1, \dots, {\cal M}_{T}$,
where ${\cal M}_{T} = \sum_{\bar{e}=-\infty}^{\infty} {\cal M}_{\bar{e}}$.
For each set $\{ {\cal M}_{\bar{e}} \}$
only a special configuration enters eq.(\ref{eq:Z-e-repr-all}).
The properties of the kernel enforce the charge neutrality condition
$\sum_{\bar{e}=-\infty}^{\infty} {\cal M}_{\bar{e}} \, \bar{e}  = 0$.

\paragraph{Coulomb gas representation II: $n$ charges}
A different Coulomb gas representation can be obtained if 
we apply the Poisson transformation to eq.(\ref{eq:Z-e-repr}) 
(or to eq.(\ref{eq:Z-Villain})). We 
obtain the $n$-representation of the QDVM
\begin{eqnarray}
\label{eq:Z-n-repr}
%--------------------------------------------------------------------
\Z{}{}(\J{}{}) &=& \Z{}{0}(\J{}{}) \cdot \Z{}{(n)}(\J{}{(n)})
\nonumber\\
\Z{}{(n)}(\J{}{(n)}) &=& \multisum{n}{\tau} 
                   \e^{-{1 \over 2} \; 
                     \sum_{\tau,\tau \pr =  0}^{N-1} \;
                   n_{\tau}  \;\; {\cal D}_{\tau \, \tau \pr} \;
                   n_{\tau \pr}
                       + \; \sum_{\tau}  \; 
                         \J{\tau}{(n)} n_{\tau} } \; ,
%--------------------------------------------------------------------
\end{eqnarray}
which is again a gas of interacting integer charges
$n_{\tau} \in ]-\infty, \infty[$. The partition function $\Z{}{0}(\J{}{})$
of the free Brownian particle~\cite{kn:Weiss-98} appears and 
again $\J{\tau}{(n)}$ can be obtained explicitly 
from $\J{\tau}{}$~\cite{kn:Falci-Weiss}.
The generating functional $\Z{}{(n)}(\J{}{(n)})$ has the
same form of eq.(\ref{eq:Z-e-repr})
with a different interaction given by 
\begin{eqnarray}
  \label{eq:kernel-n-repr}
  {\cal D}_{\tau} &=&   {\Lambda \over V} \, \deltad{\tau}{} +
                        \Lambda^2 \G{\tau}{0} = 
        \cutsum{}{} \,
          \left\{ {1 \over V} \, + \, {1 \over 
          m \omega^2  - \alpha(\omega)/2} \right\} \, \e^{-i \omega \tau}
\end{eqnarray}
where $\G{\tau}{0}$ is the discrete Green's function of the free Brownian 
particle~\cite{kn:Weiss-98}. 
%For ohmic dissipation and large $\tau$,  
%${\cal D}_{\tau} \sim (2/ \alpha) \ln \tau$. 
By proceeding as before, a Coulomb gas form for 
$\Z{}{(n)}(\J{}{(n)})$ is obtained, structurally identical to 
eq.(\ref{eq:Z-e-repr-all}).

\subsection{Self-duality}
\label{sec:Self-duality}
The fact that $\Z{}{(e)}(\J{}{(e)})$ and 
$\Z{}{(n)}(\J{}{(n)})$ are structurally identical 
indicates that the QDVM possesses a self-dual structure. The
relation between the interactions $\Delta_{\tau}$ and 
${\cal D}_{\tau}$ assumes a simple form if we introduce the 
function $h^0(w) := {Z_T(-i w ) / R_Q}$. 
Then    
$
(1 / \Lambda) \; {\cal D}_{\omega} =       
%~~~~~~
  2 \pi h^0(| \omega |)  /  | \omega |
         +  {1 / V}        =:  
%~~~~~~ 
  (2 \pi / | \omega |) \; h(| \omega |)
$
and 
$
(1 / \Lambda) \; \Delta_{\omega} \;= \;   
        (2 \pi / | \omega |) \; [1 / h(| \omega |)]
$.
So the exact self duality in the QDVM is given by
$h(\omega) \; \leftrightarrow \; {1 / h(\omega)}$,  
which is a sort of a Norton transformation.

For low frequencies we obtain 
the $\alpha \leftrightarrow 1/\alpha$ and the 
sub-ohmic $\leftrightarrow$ super-ohmic dualities analogous to the 
ones possessed by the WB model~\cite{kn:Schmid,kn:Sassetti}.

Self-duality is a powerful tool for obtaining nonperturbative 
results. 
For instance 
from our low-energy results 
we can infer the existence of the critical line
$\alpha = 1$ for ohmic dissipation in the QDVM, analogously to the 
WB model~\cite{kn:Schmid}. For any dual mapping,  
perturbative results in a certain representation may correspond to 
nonperturbative results in a different representation
(see the next section).
An appealing feature of self-duality is that one
can write exact relations 
(as the eqs. (\ref{eq:phiphi-nn}) and (\ref{eq:phiphi-ee}) 
we will use below) and exact equations for the correlation
functions~\cite{kn:Falci-Weiss}.

\section{\large{THE WB MODEL}}
\label{sec:WB}
\subsection{Mapping of the QDVM onto the WB model}
\label{sec:Mapping-onto-CL}
Now we show that if we take the continuum limit in a special way 
the QDVM on a lattice reduces exactly to the WB model. 
Indeed from the $n$-representation we
obtain the exact perturbation series in $V$ as given in ref.\cite{kn:Schmid}. 
The proof goes as follows. First 
one considers the Coulomb gas representation for $n$-charges
of the QDVM with the substitution 
$V \longrightarrow \widetilde{V} = \Lambda / 2 \ln(2 \Lambda/V)$.
Then one shows that the configurations where only the charges 
$n_l= \pm 1$ appear give the exact perturbation expansion in $V$ for the 
WB potential~\cite{kn:Schmid}.
Finally one shows that all the other charge configurations 
carry an additional factor $\Lambda$ with a negative power and give a 
vanishing contribution to 
the partition function as $\Lambda \to \infty$.

The consequence of this mapping is that the WB model 
possesses an exact self duality for arbitrary $T$ and arbitrary 
spectral densities which generalizes
the approximate Schmid self-duality~\cite{kn:Schmid}. 

Notice that the 
exact self-dual mapping is not evident if one starts directly 
from the WB model. 
In fact the perturbation expansion is recasted in a Coulomb gas 
with only $n_l = \pm 1$ charges, whereas
in the instanton expansion (the dual Coulomb gas) all kind of charges 
$e_l \in ]-\infty,\infty[$ may appear.
Self-duality can be exploited 
by carrying over the exact results for the
correlation functions of the QDVM  to the WB model.

\subsection{Impedance of a JJ circuit}
\label{sec:impedance}
As an application we calculate the leading terms of the 
retarded $\langle\varphi\varphi\rangle$ correlator which
is related to 
%the mobility of a Brownian particle in a WB 
%potential~\cite{kn:Schmid}, to 
the impedance $\mbox{\textsf Z}(\Omega)$
seen by the source of
the circuit in fig.1
%fig.\ref{fig:JJcircuit}.  
%and to the admittance of a 
%Luttinger Liquid with a barrier. 
Starting from the thermal $\langle\varphi\varphi\rangle$ correlator of 
the Villain model we perform an analytic continuation,
$\omega_n \to \,- i \omega$), 
and then take
the continuum limit with the prescription of 
sec.\ref{sec:Mapping-onto-CL}.
It would be desirable to perform the analysis in 
real-time~\cite{kn:Weiss-98}, but at present we content ourself 
to
check the QDVM method against
available exact results~\cite{kn:exact}. We
we start with the exact expression for the correlator
$(1 / \Lambda)\; \langle\varphi\varphi\rangle_{\omega_n}$ 
vs. $\langle nn \rangle_{\omega_n}$
\begin{eqnarray}
\label{eq:phiphi-nn}
        \Lambda \G{\omega_n}{0} - 
\left( \Lambda \G{\omega_n}{0} \right)^2 \; \Lambda 
        \langle nn \rangle_{\omega_n}
\; \longrightarrow \; 
x_R^{0}(\omega) - \left[ x_R^{0}(\omega) \right]^2 \;
\left[ X_R^{(n)}(0) - X_R^{(n)}(\omega) \right] \;,
\end{eqnarray}
where $x_R^{0}$ is the retarded correlator of the free
Brownian particle. and $X_R^{(n)}$ can be evaluated by analytic
continuation of the first term of the Coulomb gas expansion
\begin{eqnarray}
2 \Lambda \; \sum_{\sigma} \; \e^{ {\cal D}_{\sigma} - {\cal D}_{0}} \; 
\e^{i \omega_n \sigma} \quad
\stackrel{\omega_n \to \,- i \omega}{\mbox{{\Large{$\longrightarrow$}}}} \quad
- 4 \Lambda^2 \; \e^{-\Lambda/\widetilde{V}} \; 
\int_0^{\infty} ds \; e^{i \omega s} \; 
\Im e^{F_0^>(s)}
\nonumber
\end{eqnarray}
Here $F_0^>(s)$ accounts for free circuit 
fluctuations~\cite{kn:Ingold-Nazarov,kn:Falci-Bubanja-Schoen,kn:environment,kn:Weiss-98}
$$
{F_0^>(s)} = 
\int_0^{\infty} {d\omega \over \omega} \; 
        {{2 \, \Re Z_T(\omega)} \over R_Q} \; 
\left[ \ctgh {\beta \omega \over 2} \; ( \cos \omega s - 1) - i \sin \omega s  \right]
$$
In particular 
$
\Im X_R^{(n)}(\omega) = \left( {V / 2} \right)^2 \; \left[ 
{P_0(\omega)} - {P_0(-\omega)} 
\right] 
$, 
and its Hilbert transform gives $\Re X_R$. We introduced
$
P_0(\omega) = \int_{-\infty}^{\infty} ds \;  \exp\{i \omega s\} \; 
\exp\{F_0^>(s)\}
$
which can be calculated numerically for almost arbitrary 
environment at all 
$T$~\cite{kn:Ingold-Nazarov,kn:environment,kn:Falci-Bubanja-Schoen}

As a check of the theory we calculated analytically the low-frequency 
behavior of $\mbox{\textsf Z}(\Omega)$ 
at $T=0$, using the toy ohmic environment
$
\Re Z_T(\omega)/R_Q = 
  \alpha^{-1} \; \exp\{-\mid \omega \mid / \bar{\omega} \}
$. For $\alpha < 1 $ we find 
\begin{equation}
\label{eq:impedance-pert}
{\Re \, \mbox{\textsf Z}(\Omega) \over R_Q } = 
        {\Omega \over 2 \pi} \; \Im x_R(-\Omega) 
\;
\stackrel{\Omega \to 0^+}{\longrightarrow} \;
{{\Re Z_T(\Omega)} \over R_Q } \; \left\{1 -  
         A \; \Omega^{-2 + 2 / {\alpha}} 
         -  B   \Omega^2 \right\} 
\end{equation}
which reproduces known exact results~\cite{kn:exact}, 
i.e. the  leading behavior of the correlation function is 
        $\sim \Omega^2$ for $0 < \alpha < 1/2$ and
        $\sim \Omega^{-2 + 2 / \alpha}$
        for $1/2 < \alpha < 1$.
It corresponds to a vanishing junction {\em admittance},
%%%%%$Y_J(\omega) = \dots \approx \dots$ 
which
describes the Coulomb blockade.
The coefficients in (\ref{eq:impedance-pert}) depend on the details of
the environment at frequencies of the order of the cutoff $\bar{\omega}$ 
and can be evaluated numerically.

Then we start with the exact expression of the  
$\langle\varphi\varphi\rangle$ correlation function
in the $e$-representation 
and then perform the analytic continuation
\begin{eqnarray}
\label{eq:phiphi-ee}
{1 \over \Lambda} \; \; \langle\varphi\varphi\rangle_{\omega_n} 
&=& \Lambda \G{\omega_n}{\widetilde{V}} - 
\left( {2 \pi \widetilde{V} \Lambda \G{\omega_n}{\widetilde{V}} 
	\over \omega_n} \right)^2 
\Lambda \langle ee \rangle_{\omega_n}
\quad
\stackrel{\omega_n \to \,- i \omega}{\mbox{{\Large{$\longrightarrow$}}}} \quad
\nonumber \\
&&
x_R^{\widetilde{V}}(\omega) - 
\left[ { 2 \pi \widetilde{V} \, x_R^{\widetilde{V}}(\omega) 
	    \over \omega} \right]^2 
\left[ X_R^{(e)}(0) - X_R^{(e)}(\omega) \right]
\end{eqnarray}
where $\G{\omega_n}{\widetilde{V}}$ is the discrete Green's function and 
$x_R^{\widetilde{V}}$ 
is the retarded correlation function for the damped
oscillator.
As before we consider the first contribution in the 
Coulomb gas expansion eq.(\ref{eq:Z-e-repr-all}) for 
$\Lambda \langle ee \rangle_{\omega_n}$
\begin{eqnarray}
\label{eq:retarded-ee}
2 \Lambda \; \sum_{\sigma} \; 
\e^{ {\Delta}_{\sigma} - {\Delta}_{0}} \; 
\e^{i \omega_n \sigma} \quad
\stackrel{\omega_n \to \,- i \omega}{\mbox{{\Large{$\longrightarrow$}}}} \quad
- 4 \Lambda^2 \;
\int_0^{\infty} ds \; e^{i \omega s} \; 
\Im e^{F^>(s)}
\end{eqnarray}
A different correlator for the fluctuations appears, $F^>(s)$, coming
from the analytic continuation of $\Delta_{\tau}$, from which 
we can define a $P(\omega)$ function which enters
the expression for $\mbox{\textsf Z}(\Omega)$. 
To evaluate $\mbox{\textsf Z}(\Omega)$ at all frequencies we need a 
regularization scheme, which is not needed for checking 
analytically 
the power-low behavior at low frequencies. For $\alpha > 1$ we find
\begin{eqnarray}
{\Re \, \mbox{\textsf Z}(\Omega) \over R_Q } \;\;
\stackrel{\Omega \to 0^+}{\longrightarrow } \;\;
 A \; \Omega^{2   
{\alpha} - 2}  + B \;\Omega^{2}
\nonumber
\end{eqnarray}
which in this case gives also the junction {\em impedance}. Again 
known exact results~\cite{kn:exact} are reproduced.

\section{\large{CONCLUSIONS}}
\label{sec:conclusions}
In summary we have exploited the 
Schmid~\cite{kn:Schmid} self-duality in the WB model.
Using the QDVM we have shown that it is exact, that it holds 
for finite 
temperatures and that it can be extended to 
an environment with general frequency dependence
of the spectral density. 
Exact low-frequency results 
for the strictly ohmic environment are reproduced using this method.

The study of a general environment is important 
for design and performances 
of the devices, for instance 
for an optimal control of the noise.
In the Coulomb blockade regime
it is possible to treat a 
general electromagnetic environment  in realistic situations
using the $P$-function
theory~\cite{kn:Ingold-Nazarov,kn:Falci-Bubanja-Schoen,kn:environment}.
The QDVM theory can be
combined with the $P$-function
theory to calculate quantities of direct experimental interest
different regimes.

\baselineskip=15pt

\vspace{5mm}
\paragraph{Acknowledgments} G.F. acknowledges G. Sch\"on for motivating 
the last part of this work, R. Fazio and M. Sassetti for discussions and 
suggestions, Universit\"at Stuttgart (Germany) for hospitality, 
EU (FMRX CT 960042) and INFM (PRA-QTMD) for support. 
This work is dedicated to Prof. A. Schmid who recently passed away.

\newpage
\noindent
{\large \bf Figure captions}

\vspace{10mm}
\noindent
{\bf Figure 1}: The JJ circuit described by the WB model: the special
variable is the phase across the junction $\varphi$ and  
the impedance $Z(\omega)$ is modeled by an environment of quantum
oscillators.
Classical and quantum fluctuations of $\varphi$ are modulated by the 
fluctuations of the environment.

\vspace{10mm}
\noindent
{\bf Figure 2}: (a) A typical trajectory
in imaginary time for the coordinate of 
the QDVM, $ \varphi(\tau) - 2 \pi m(\tau)$. 
(b) A typical configuration after the 
elimination of $\{\varphi_{\tau} \}$: the set $\{m_{\tau} \}$ is 
associated to a one-dimensional interface; each elementary 
($1/\Lambda$) piece of surface interacts with all the other 
elements.
(c) The corresponding charge configuration in the Coulomb gas
representation eq.(\ref{eq:Z-e-repr-all}): in general charges 
may take any integer value in $e_l \in ]-\infty,\infty[$.

\newpage
\noindent
{\large \bf Falci and Weiss - Figure 1}

\vspace{30mm}
\noindent

\begin{figure}[h!]
\label{fig:JJcircuit}
\begin{center}
\resizebox{13cm}{!}{\includegraphics{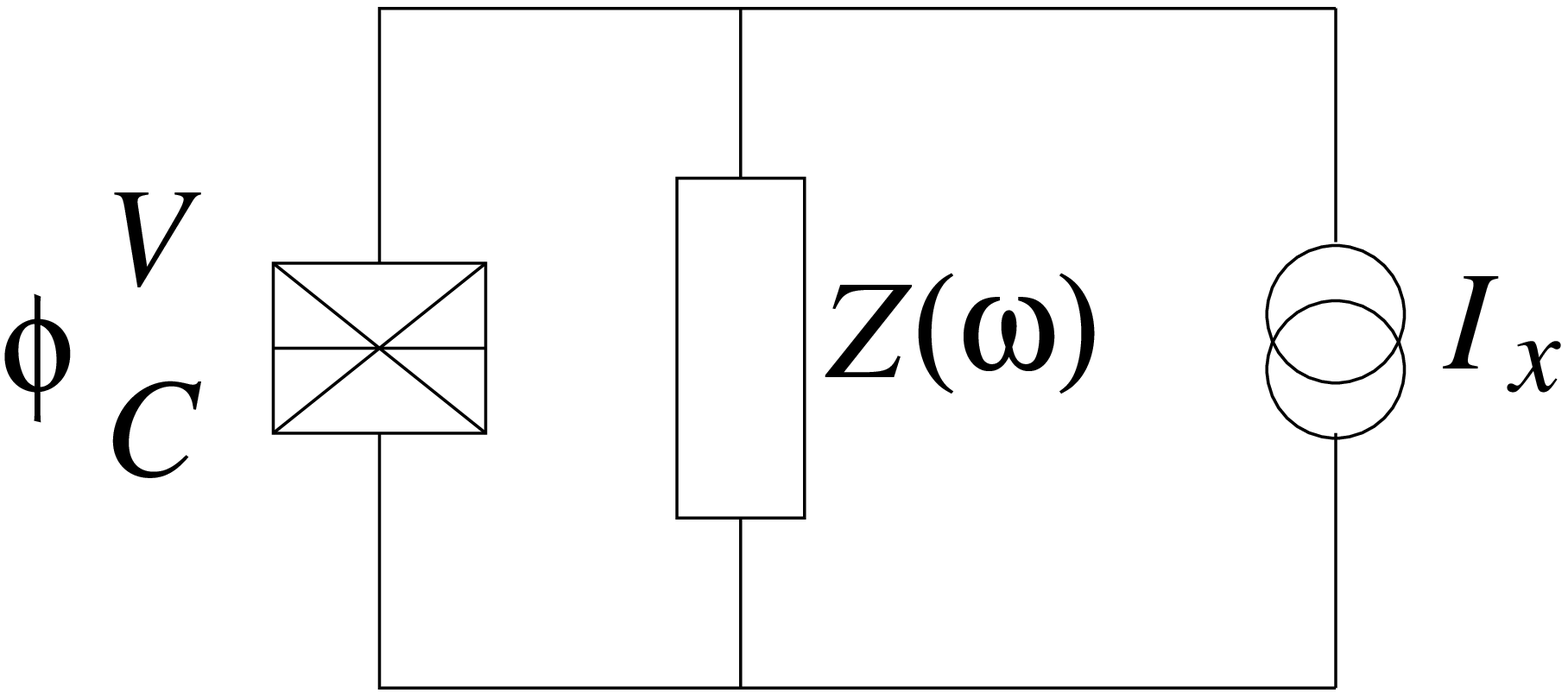}} 
\end{center}
\end{figure}

\newpage
\noindent
{\large \bf Falci and Weiss - Figure 2}

\vspace{30mm}
\noindent
\begin{figure}[h!]
\label{fig:discretex}
\begin{center}
\resizebox{13cm}{!}{\includegraphics{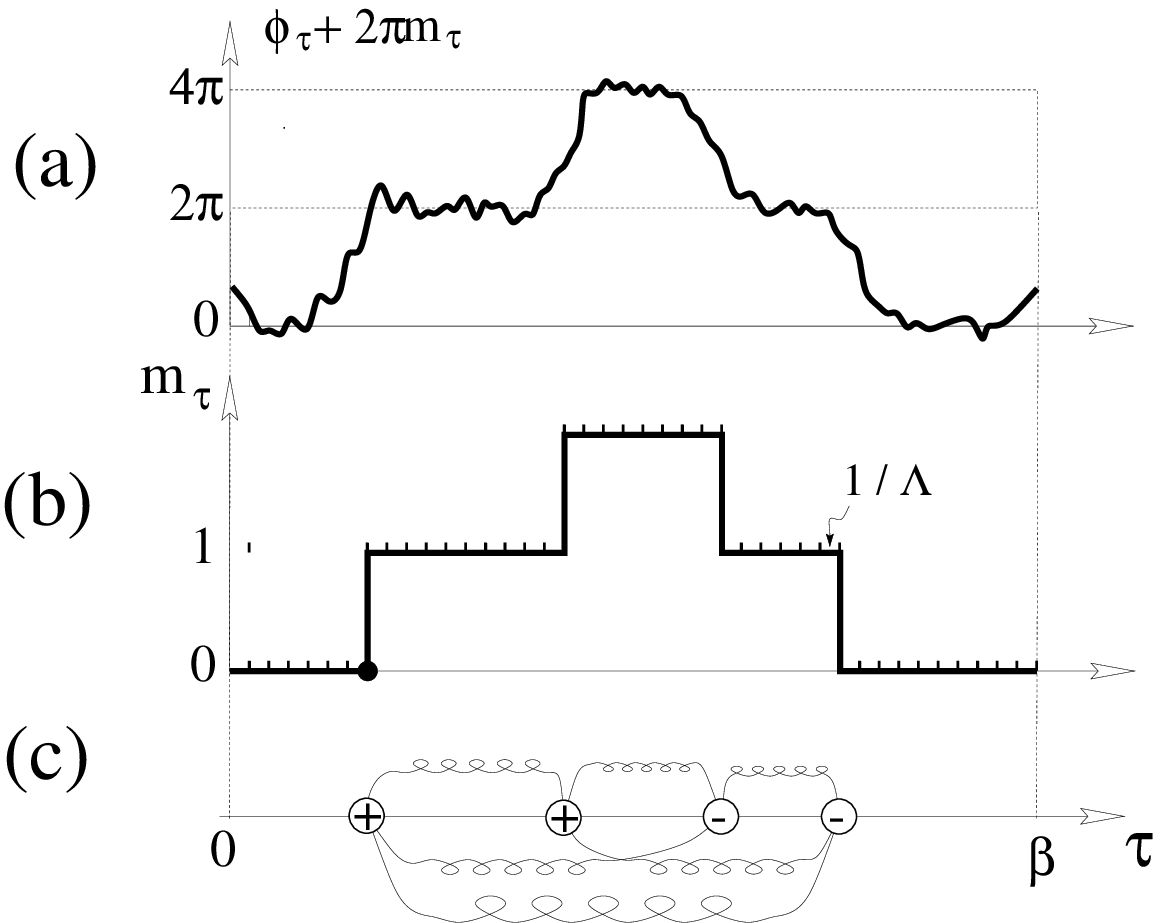}} 
\end{center}
\end{figure}

\begin{thebibliography}{99}
\bibitem{kn:Weiss-98} U. Weiss, {\em Quantum Dissipative Systems}, 
  Series in Modern Condensed Matter Physics, vol. 2, second Edition,
  World Scientific Singapore 1998.
%\bibitem{kn:Savit-80} Savit, Rev. Mod. Phys.
%\bibitem{kn:Weiss-Grabert} U. Weiss H. Grabert
\bibitem{kn:Leggett} A.J. Leggett, in ``Directions in Condensed Matter 
	Physics'', vol 1., G. Grinstein and G. Mazenko Eds., 
	World Scientific, Singapore, 1986, p. 187.
\bibitem{kn:Caldeira-Leggett} A.O. Caldeira and A.J. Leggett, Phys. Rev. Lett.
			{\bf 46}, 211 (1981)
\bibitem{kn:revs}  G. Sch\"on and A.D. Zaikin, Phys. Rept. 
		{\bf 198}, 237 (1990);
                  D.A. Averin and K.K. Likharev, in 
	{\em Quantum Effects in Small Disordered Systems}, B.L. Altshuler,
	P.A. Lee and R.A. Webb Eds., Elsevier, Amsterdam, 1991. 
%\bibitem{kn:Kane-Fisher} C. Kane and M.P.A. Fisher
%\bibitem{kn:Fisher} Fisher ??
\bibitem{kn:Ingold-Nazarov} G.L. Ingold and Yu.V. Nazarov, in 
	{\em Single Charge Tunneling}, H. Grabert and M. Devoret Eds., 
	Plenum, New York, 1991. 
\bibitem{kn:Leggett-84} A.J. Leggett, Phys. Rev. B {\bf 30}, 1208 (1984).
\bibitem{kn:Falci-Bubanja-Schoen} G. Falci, V. Bubanja and G. Sch\"on , 
  Europhys. Lett. {\bf 16}, 109 (1991); Zeit. Phys. B {\bf 85}, , 451 (1991).
\bibitem{kn:Schmid} A. Schmid, Phys. Rev. Lett. {\bf 51}, 1506 (1983).
\bibitem{kn:Fisher-Zwerger} M.P.A. Fisher and W. Zwerger 
	Phys. Rev. B {\bf 32}, 6190 (1985); W. Zwerger,
	Phys. Rev. B {\bf 35}, 4737 (1987)
\bibitem{kn:Sassetti} M. Sassetti, H. Schomerus and U. Weiss,
	Phys. Rev. B {\bf 53}, R2914 (1996).
\bibitem{kn:Fendley-Saleur} P. Fendley and H. Saleur, 
	cond-mat/9804173, 16 Apr. 1998
\bibitem{kn:exact} P. Fendley, A.W.W. Ludwig and H. Saleur,
	Phys. Rev. Lett. {\bf 75}, 8934 (1995); F. Lesage, H. Saleur
	and S. Shorik,  Phys. Rev. Lett. {\bf 76}, 3384 (1996);
	U. Weiss, Sol. St. Comm. {\bf 100}, 281 (1996); 
	U. Weiss, in
	{\em Tunneling and its implications}, 
	L.S. Schulman, A. Ranfagni and D. Mugnai Eds., 
	Word Scientific, Singapore 1997.
\bibitem{kn:Chui-Weeks-82} Chui and Weeks, Phys. Rev. B {\bf 14}, 4978 (1983).
\bibitem{kn:Villain} J. Villain, J. Phys. (Paris) {\bf 36}, 581 (1976)
\bibitem{kn:arrays} 
	S. Korshunov, Europhys. Lett. {\bf 9}, 107 (1989); 
	 W. Zwerger, Z. Phys. B {\bf 78}, 111 (1990);
	R. Fazio and G. Sch\"on, Phys. Rev. B {\bf 43}, 5307 (1991).
\bibitem{kn:Falci-Weiss} G. Falci and U. Weiss, preprint 1998.
\bibitem{kn:environment} M.H. Devoret, D. Esteve, H. Grabert, G.L. Ingold,
	H. Pothier and C. Urbina, Phys. Rev. Lett. {\bf 64}, 1824 (1990); 
	S.M. Girvin, L. Glazman, M. Jonson, D.R. Penn and M.D. Stiles 
	Phys. Rev. Lett. {\bf 54}, 3183 (1990); 
	D.A. Averin, Yu.V. Nazarov and A.A. Odintsov, Physica B {\bf 165\&166},
	945 (1990); 
	Yu.V. Nazarov, J. Siewert, G. Falci, 
	Europhys. Lett. {\bf 38}, 365 (1997); 
        R. Cristiano, L. Frunzio, G. Falci and A. Tagliacozzo, in
	{\em Tunneling and its implications}, 
	L.S. Schulman, A. Ranfagni and D. Mugnai Eds., 
	Word Scientific, Singapore 1997, 
	p. 161.
\end{thebibliography}
\end{document}